\documentclass[journal,twoside,web]{ieeecolor}
\usepackage{tmi}
\usepackage{cite}
\usepackage{amsmath,amssymb,amsfonts}
\usepackage{graphicx}
\usepackage{textcomp}
\usepackage{bbm}
\usepackage{multirow}
\usepackage{lineno} 
\usepackage{hyperref}
\usepackage{algorithm,algpseudocode}
\usepackage{tablefootnote}
\usepackage{threeparttable}
\usepackage{makecell}
\algnewcommand{\Inputs}[1]{%
  \State \textbf{INPUT:}
  \Statex \hspace*{\algorithmicindent}\parbox[t]{.8\linewidth}{\raggedright #1}
}
\algnewcommand{\Initialize}[1]{%
  \State \textbf{Initialize:}
  \Statex \hspace*{\algorithmicindent}\parbox[t]{.8\linewidth}{\raggedright #1}
}
\algnewcommand{\ALGORITHM}[1]{%
  \State \textbf{ALGORITHM:}
  \Statex \hspace*{\algorithmicindent}\parbox[t]{.8\linewidth}{\raggedright #1}
}
\algnewcommand{\OUTPUT}[1]{%
  \State \textbf{OUTPUT:}
  \Statex \hspace*{\algorithmicindent}\parbox[t]{.8\linewidth}{\raggedright #1}
}
\def\BibTeX{{\rm B\kern-.05em{\sc i\kern-.025em b}\kern-.08em
    T\kern-.1667em\lower.7ex\hbox{E}\kern-.125emX}}
\begin{document}
\switchlinenumbers
\title{Score-based Diffusion Models with Self-supervised Learning for Accelerated 3D Multi-contrast Cardiac MR Imaging}
\author{Yuanyuan Liu,  Zhuo-Xu Cui, Shucong Qin, Congcong Liu,  Hairong Zheng,\IEEEmembership{Senior Member, IEEE},  Haifeng Wang,\IEEEmembership{Senior Member, IEEE},  Yihang Zhou, Dong Liang,\IEEEmembership{Senior Member, IEEE}, and Yanjie Zhu,\IEEEmembership{Member, IEEE} 
\thanks{This study was supported in part by the National Key R\&D Program of China nos . 2021YFF0501402, 2020YFA0712200, National Natural Science Foundation of China under grant nos. 62322119, 62201561, 62206273, 62476268, 12226008, 62125111, 62106252, 81971611, and 81901736, the Guangdong Basic and Applied Basic Research Foundation under grant no. 2021A1515110540, and the Shenzhen Science and Technology Program under grant no. RCYX20210609104444089.(Corresponding author: Yanjie Zhu) }
\thanks{Yuanyuan Liu, Shucong Qin, Congcong Liu, Hairong Zheng, Haifeng Wang and Yanjie Zhu are with Paul C. Lauterbur Research Center for Biomedical Imaging, Shenzhen Institute of Advanced Technology, Chinese Academy of Sciences, Shenzhen, Guangdong, China (e-mail:$\left \{\text{liuyy; cc.liu; hr.zheng; hf.wang1; yj.zhu}\right \}$@siat.ac.cn;idqinshucong\\@163.com).}
\thanks{Zhuo-Xu Cui, Yihang Zhou and Dong Liang are with Research Center for Medical AI, Shenzhen Institute of Advanced Technology, Chinese Academy of Sciences, Shenzhen, Guangdong, China (e-mail:  $\left \{\text{zx.cui; yh.zhou2, dong.liang}\right \}$@siat.ac.cn). }
\thanks{Congcong Liu is also with Shenzhen College of Advanced Technology, University of Chinese Academy of Sciences.}
\thanks{Yuanyuan Liu and Zhuo-Xu Cui contributed equally to this study.} }

\maketitle
\begin{abstract}
Long scan time significantly hinders the widespread applications of  three-dimensional multi-contrast cardiac magnetic resonance (3D-MC-CMR) imaging.  This study aims to accelerate 3D-MC-CMR acquisition by a novel method based on score-based diffusion models with self-supervised learning. 
Specifically, we first establish a mapping between the undersampled k-space measurements and the MR images, utilizing a self-supervised Bayesian reconstruction network.  Secondly, we develop a joint score-based diffusion model on 3D-MC-CMR images to capture their inherent distribution. The 3D-MC-CMR images are finally reconstructed using the conditioned Langenvin Markov chain Monte Carlo sampling. This approach enables accurate reconstruction without fully sampled training data. Its performance was tested on the dataset acquired by a 3D joint myocardial $\text T_{1}$ and $\text T_{1\rho}$ mapping sequence. The $\text T_1$ and $\text T_{1\rho}$ maps were estimated via a dictionary matching method from the reconstructed images.
 Experimental results show that the proposed method outperforms traditional compressed sensing and existing self-supervised deep learning MRI reconstruction methods.  It also achieves high quality $\text T_1$ and $\text T_{1\rho}$ parametric maps close to the reference maps, even at a high acceleration rate of 14. 
\end{abstract}

\begin{IEEEkeywords}
3D cardiac magnetic resonance imaging, self-supervised, diffusion models, multi-contrast
\end{IEEEkeywords}

\section{Introduction}
\label{sec:introduction}
\IEEEPARstart{M}{yocardial} parameter mapping, including $\text T_{1} $, $\text T_{1\rho}$, $\text T_{2}$, and $\text T_{2}^*$ mapping, is an important technique in cardiovascular MR imaging (CMR). It enables the quantification of MR relaxation times, serving as biomarkers for a range of cardiac pathologies\cite{rao2022myocardial}. Previous studies have shown that inflammation, fibrosis, and amyloid deposition can lead to increased native $\text T_{1} $ values\cite{haaf2016cardiac}, while conditions like iron deposition or significant fat accumulation can result in decreased $\text T_{1} $ values\cite{meloni2021myocardial}. Elevated $\text T_{1\rho}$ has been observed in both acute and chronic myocardial infarction\cite{bustin2023magnetic}, as well as in nonischemic myocardial diseases, including dilated and hypertrophic cardiomyopathy\cite{thompson2021endogenous}.  In contemporary CMR, there is a growing emphasis on  characterizing multiple relaxation times simultaneously through the acquisition of multi-contrast (MC) images.  
Examples include joint myocardial $\text T_{1} $ and $\text T_{2}$ mapping \cite{akccakaya2016joint}, cardiac MR fingerprinting (MRF)\cite{velasco2022MRF}, and CMR multitasking\cite{MTP2018}. $\text T_{1}$ and $\text T_{1\rho}$ parameter maps provide complementary information about the myocardium, and combining these two maps may enhance diagnostic sensitivity and increase confidence in diagnosing suspected cardiomyopathy.
 Furthermore, 3D-CMR can provide information of whole-hearts relative to
2D imaging with limited coverage, but it suffers from the long scan time, especially for multi-parameter mapping \cite{cruz20203}. Therefore, accelerated 3D multi-contrast imaging of CMR is highly desirable for practical use.  
\par One major strategy to accelerate 3D-MC-CMR is undersampling the $k$-space data and subsequently reconstructing images from this undersampled data using regularized techniques with hand-crafted priors. Various priors have been employed in 3D-MC-CMR reconstruction, such as subspace modeling\cite{zhao2015accelerated}, patch-based\cite{milotta20203d,zhang2015accelerating}, and tensor-based low-rankness \cite{phair2023free}. These approaches can achieve high acceleration rates while maintaining high reconstruction quality. However, constructing priors is non-trivial, and the iterative nature of these methods leads to time-consuming computations. Deep learning (DL) has emerged as a successful tool in MR reconstruction, reducing computational complexity and enhancing reconstruction quality.  However, most DL-based reconstruction methods rely on supervised learning that requires extensive fully sampled datasets for training. Acquiring  high-quality, fully sampled $k$-space data  for 3D-MC-CMR  is  challenging due to the  extreme long scan time. Heart rate fluctuations and respiratory variations during the prolonged scan time of the 3D-MC-CMR imaging can lead to misalignment of the cardiac phase and instability in respiratory motion correction.  Despite the use of cardiac triggering and respiratory navigation, these instabilities may still result in scan failures or compromised image quality.  Therefore, these supervised learning methods are unsuitable for accelerating 3D-MC-CMR.

\par To address this issue, self-supervised learning methods have been adopted. One strategy directly reconstructs images from the undersampled $k$-space data by enforcing the data consistency using the physical model in the reconstruction \cite{martin2021physics,liu2021magnetic}.
 Another strategy is training the networks with a sub-sample of the undersampled $k$-space data acquired.
A notable example is the self-supervised learning via data undersampling (SSDU) method \cite{yaman2020self}. SSDU divides the measured data into two distinct subsets: one for use in the data consistency term of the network, and the other to compute the loss function. Consequently, network training can be completed  using only undersampled data. Based on SSDU, the reference \cite{zhou2022dual} introduces a dual-domain self-supervised (DDSS) method which trains the network in both $k$-space and image domain and applies it in  accelerated non-cartesian MRI reconstruction. 

%

Recently, the score-based diffusion model \cite{song2021maximum,chung2022score} has been applied to MR reconstruction and has shown promising results in reconstruction accuracy and generalization ability. It learns data distribution rather than end-to-end mapping using the network between undersampled $k$-space data and images, and therefore is recognized as an unsupervised learning method for solving inverse problems of MR reconstruction. Specifically, it perturbs the MR  data distribution to a tractable distribution (i.e. isotropic Gaussian distribution) by gradually injecting Gaussian noise, and one can train a neural network to estimate the gradient of the log data distribution (i.e. score function). Then the MR reconstruction can be accomplished by sampling from the data distribution conditioned on the acquired $k$-space measurement. Although treated as an unsupervised technique \cite{song2021}, score-based reconstruction still needs amounts of high-quality training images to estimate the score function. The image collection is also challenging for 3D-MC-CMR. A simple and straightforward strategy to solve this issue is using the reconstruction results of traditional regularized or self-supervised learning methods as training samples. However, the residual reconstruction errors may lead to discrepancy in the data distribution of diffusion models. Previous studies have shown that Bayesian neural networks predict the target with uncertainty by training the networks with random weights,  provide a more comprehensive sampling over the whole distribution and may avoid the discrepancy\cite{jospin2022hands}. 
\par In this study, we sought to develop a self-supervised reconstruction method for 3D-MC-CMR using the score-based model with a Bayesian convolutional neural network (BCNN). Specifically, a self-supervised BCNN was developed based on the neural network driven by SPIRiT-POCS model \cite{cui2023equilibrated}. Then the score-based model was trained by the samples over the network parameter distribution of BCNN. To better utilize the correlations among MC images, we used a joint probability distribution of MC data instead of the independent distribution in the original score-based model. The proposed method (referred to as "SSJDM") was tested using the data acquired by a 3D joint myocardial $\text T_{1}$ and $\text T_{1\rho}$ mapping sequence based on the pulse sequence used for joint $\text T_1$ and $\text T_2$ mapping \cite{akccakaya2016joint,henningsson2022cartesian}. In particular, the readout process was the same as that in \cite{akccakaya2016joint,henningsson2022cartesian}, except that the $\text T_{1}$ and $\text T_2$  preparation pulse was replaced by a spin-lock and inversion preparation pulse\cite{yang2023robust}.  The SSJDM method outperformed the state-of-the-art methods, including regularized and self-supervised DL reconstructions, at a high acceleration rate (R) of up to 14. The obtained $\text T_{1}$ and $\text T_{1\rho}$ values were also comparable with the reference values.

\section{Related Work}
\subsection{Self-supervised Learning in MRI Reconstruction}
\par Self-supervised learning leverages only the measurement data itself for training, without requiring the fully-sampled data. In MRI reconstruction, self-supervised learning has been a long-standing topic of interest. For example, the generalized autocalibrating partially parallel acquisition (GRAPPA) uses auto-calibration signal (ACS) data from the central region of $k$-space to train linear interpolation kernels, which are applied to the measurement data itself to recover missing $k$-space data\cite{griswold2002generalized}. With the development of deep learning, linear estimations of GRAPPA kernels are replaced by non-linear estimations using a scan-specific convolutional neural network (CNN)\cite{akccakaya2019scan,kim2019loraki}. This CNN is trained using ACS data without requiring external training databases, achieving more accurate interpolation. However, relying solely on ACS data for self-supervised training can result in the loss of high-frequency information.
To enhance reconstruction performance, one effective approach is to integrate physical models into the network as regularizations during training. These models, such as coil sensitivity and MR relaxometry, enforce the network to be more consistent with the principles of MR imaging physics, thereby improving its performance \cite{martin2021physics,liu2021magnetic}.
 Another approach attempts to train networks by splitting the measurements into two disjoint sets, such as the SSDU \cite{yaman2020self} and DDSS methods \cite{zhou2022dual}.
Additionally, researchers have investigated the interpretability of self-supervised learning. Luo et al. developed a Hankel low-rank model-driven CNN by leveraging the convolutional relationship between structured low-rank Hankel matrices and null-space filters\cite{luo2023matrix}. Moreover, Gan et al. proved the convergence of deep unfolding and provided theoretical guarantees for self-supervised learning\cite{gan2023self}. 
\subsection{Generative Models in MRI}
\subsubsection{Early Generative Models}Since 2000s, generative models, particularly Generative Adversarial Networks (GAN)\cite{goodfellow2014generative}, have achieved remarkable success and have produced high-quality results in several self-supervised MRI applications, such as reconstruction, image synthesis, denoising, segmentation, and super-resolution\cite{cole2020unsupervised, korkmaz2022unsupervised, yurt2022semi,zhang2023multi,ran2019denoising}.
Another widely used model is the Variational Autoencoders (VAE)\cite{kingma2013auto}, which can also generate high-quality samples. However, the one-step encode-decode process of VAE often results in blurry images due to the lack of high-frequency information, while GAN, relying purely on data-driven generators, is prone to mode collapse. Therefore, both models show limited applications in MRI reconstruction.
\subsubsection{Diffusion Generative models}
In recent years, denoising diffusion probabilistic model (DDPM) has been proposed\cite{ho2020denoising}. Compared to GAN and VAE, DDPM generates high-fidelity samples thanks to its gradual noise removal process and more stable training. Additionally, when the diffusion process is treated as continuous, transforming data into a noise distribution can be modeled with a continuous-time stochastic differential equation (SDE)\cite{song2020score}. By learning the data distribution through a score function and solving a reverse-time SDE, a score-based diffusion model is obtained.
\subsubsection{Diffusion Models in MRI Reconstruction} Given the powerful generative capabilities of diffusion models, they have been rapidly adopted for MRI reconstruction. Specifically, in the inference phase of the model, the measurement data can be treated as a condition to guide the generation of high-quality MR images  using the Bayesian formula.
For example, Chung et al.  integrated the MRI data consistency term into the reverse SDE decoding process,  guiding  generated  images to satisfy data 
consistency\cite{chung2022score}. Wu et al. addressed challenges of unstable score-based generative model training by integrating an adaptive wavelet sub-network, applying wavelet regularization to guide Langevin dynamics sampling \cite{wu2023wavelet}. This enables handling undersampled and noisy input data, significantly improving reconstruction performance. Güngöra et al. developed an unconditional diffusion prior for high-fidelity image generation and utilize this prior during inference to enhance both performance and reliability under domain shifts \cite{gungor2023adaptive}. Mardani et al. introduced a variational approach using a rigorous maximum-likelihood framework that addresses the limitations of posterior score approximation in image reconstruction. They formulated the sampling process as a stochastic optimization problem, allowing the use of off-the-shelf optimizers for faster and more adjustable sampling \cite{mardani2023variational}.
\par Another way for diffusion model-based MRI reconstruction is refining the forward and reverse processes of the diffusion model based on the physical properties of MRI, aligning it more closely with the imaging physics of MRI. For example, Cao et al. observed that in MRI reconstruction, low frequency region of $k$-space is often fully sampled, with reconstruction focusing on recovering missing high-frequency regions\cite{cao2024high}. They proposed a high-frequency diffusion model, restricting the diffusion process to high-frequency space. Guan et al. further coupled multiple methods for extracting high-frequency signals, and proposed the correlated and multi-frequency diffusion model to enhance MRI reconstruction accuracy in highly undersampled images by combining high-frequency operators to form a multi-frequency prior, reducing noise and accelerating convergence\cite{guan2024correlated}. Cui et al.incorporated physical priors from $k$-space data to constrain diffusion models, achieving more controlled generation. By interpolating high-frequency (HF) $k$-space data from low-frequency (LF) data and connecting this interpolation with the reverse heat diffusion process, they developed a model to generate missing HF data with improved accuracy\cite{cui2024physics}.
\subsubsection{Diffusion Models in Other MR Applications} Beyond MRI reconstruction, diffusion models have also shown impressive results in other MRI applications.  Levac et al. used score-based models for joint image reconstruction and motion estimation with retrospectively sub-sampled data corrupted by simulated rigid motion\cite{levac2023accelerated}. Liu et al. proposed a $k$-t self-consistency diffusion model for dynamic imaging by  designing a SDE in line with $k$-t self-consistency of the dynamic data\cite{liu2024kt}. Wang et al. implemented diffusion model for $\text T_1$ quantification in the brain, framing the estimation of quantitative maps as a conditional generation task\cite{wang2024qmri}.
\subsection{Self-supervised Diffusion Models in MRI Reconstruction}
Although diffusion models do not require paired training data, they still need fully-sampled high-quality images or $k$-space data as training datasets, which can be difficult to obtain in some practical situations. Therefore, there is a growing interest in developing diffusion models that do not require fully sampled data for training. Korkmaz et al. proposed the self-supervised diffusion reconstruction (SSDiffRecon) method \cite{korkmaz2022unsupervised}, which formulates the conditional diffusion process as an unrolled architecture, interleaving cross-attention transformers for reverse diffusion steps with data-consistency blocks for physics-driven processing. Unlike SSDiffRecon, the proposed SSJDM method introduces stochasticity through the BCNN network in the generated training data of the diffusion model to mitigate the effects of extreme generated data, thereby enhancing the overall robustness of network training. Additionally, while SSDiffRecon is designed just for single-contrast imaging, SSJDM enables multi-contrast imaging. 

\section{Material and Methods}
\subsection{Background}
\par Suppose the MR image, denoted as $\boldsymbol{x}$, is a random sample from a distribution $p(\boldsymbol{x})$, i.e.,  $\boldsymbol{x}\sim p(\boldsymbol{x})$. To sample $\boldsymbol{x}$ from $p(\boldsymbol{x})$, we can use the Langevin Markov Chain Monte Carlo (MCMC) sampling method in practical use\cite{hastings1970monte}.
In accelerated MRI, only the undersampled $k$-space data, denoted as $\boldsymbol{y}$, can be obtained and therefore accessing the distribution $p(\boldsymbol{x})$ is challenging. Fortunately, we can assume the existence of a mapping from the measurements $\boldsymbol{y}$ to the image $\boldsymbol{x}$, expressed as $\boldsymbol{x} = \boldsymbol{f_{\theta}}(\boldsymbol{y}) + n$. $\boldsymbol{f_{\theta}}$ represents a mapping parameterized by $\boldsymbol{\theta}$, where $\boldsymbol{\theta}$ follows a distribution $p(\boldsymbol{\theta})$, i.e., $\boldsymbol{\theta}\sim p(\boldsymbol{\theta})$, and $n$ denotes Gaussian noise.
Statistically,  $p(\boldsymbol{\theta})$ can be approximated by a distribution $q(\boldsymbol{\theta})$ through variational inference accomplished using BCNN. Then by substituting the learned $q(\boldsymbol{\theta})$ into the mapping mentioned above, we can employ the score-matching technique to estimate the score of $p(\boldsymbol{x})$, i.e. the gradient of $\log p(\boldsymbol{x})$, allowing the implementation of Langevin MCMC sampling.
\subsubsection{Bayesian Convolutional Neural Network}
BCNN is a type of neural network that combines traditional CNN architectures with Bayesian methods for uncertainty estimation, belonging to stochastic artificial neural networks \cite{jospin2022hands}. Specifically, it treats each weight parameter in CNNs as a random variable following a prior distribution, typically Gaussian distribution. BCNN learns the distributions of the weights, which can be used to quantify the uncertainty associated with the probabilistic predictions.
\par To design a BCNN, the first step is choosing a CNN $\boldsymbol{f_{\theta}}$ as the functional model, with its weights represented as  $\boldsymbol{\theta} $, and establishing a prior distribution for the model's parameters, denoted as  $p(\boldsymbol{\theta})$. This process can be formulated as follows: 
\begin{equation}
\label{MRI model}
\begin{aligned}
& \boldsymbol{x}=\boldsymbol{f}_{\boldsymbol{\theta}}(\boldsymbol{y})+\boldsymbol{\epsilon}, & \boldsymbol{\theta} \sim p(\boldsymbol{\theta}) 
\end{aligned}
\end{equation}
where $\boldsymbol{\epsilon}$ represents random noise with scale $\gamma$. 
Given the training data $\mathcal{D}:=\{\boldsymbol{x}, \boldsymbol{y}\}$,  we can estimate the posterior distribution of the model’s parameters $p(\boldsymbol{\theta}|\mathcal{D} )$ by Bayesian inference. Since the true posterior is often intractable, an approximate distribution $q(\boldsymbol{\theta})$ is obtained by minimizing the following the Kullback-Leibler divergence (KL-divergence) \cite{kullback1951information,candes2006robust}:
\begin{equation}
\label{min KL}
\underset{\boldsymbol{\theta}}{\operatorname{minimize}} \operatorname{KL}\left[q(\boldsymbol{\theta}) \  \|\ p(\boldsymbol{\theta} |\mathcal{D})\right]
\end{equation}
According to \cite{graves2011practical},  $q(\boldsymbol{\theta})$ can be represented as a Gaussian distribution for each individual weight ${\theta}_{s} $  as:
\begin{equation}
q(\boldsymbol{\theta}| \boldsymbol{\mu_\theta}, \boldsymbol{\sigma_\theta})=\prod_{s} \mathcal{N}({\theta}_{s} | \mu_{s}, \sigma_s)
\end{equation}
where $\boldsymbol{\theta}=\{\theta_s\}$, $\boldsymbol{\mu_{\theta}}=\{\mu_s\}$, $\boldsymbol{\sigma_{\theta}}=\{\sigma_s\}$.
However, the minimization of the KL-divergence in (\ref{min KL}) is still difficult to find for general distributions.  
Empirically, the prior distribution $p(\boldsymbol{\theta})$ can be selected as a Gaussian distribution with zero mean and standard deviation $\bar \sigma$. Then (\ref{min KL}) can be computed as\cite{bishop2006pattern}:

\begin{equation}
\label{loss functon1}
\begin{aligned}
& \min _{\boldsymbol{\mu}_{\boldsymbol{\theta}}, \boldsymbol{\sigma}_{\boldsymbol{\theta}}} \frac{1}{2 \gamma^2} \mathbb{E}_{q\left(\boldsymbol{\theta} | \boldsymbol{\mu}_{\boldsymbol{\theta}}, \boldsymbol{\sigma}_{\boldsymbol{\theta}}\right)} \log p(\mathcal{D} | \boldsymbol{\theta}) \\
& +\frac{1}{2 \bar{\sigma}^2}(\left\|\boldsymbol{\mu}_{\boldsymbol{\theta}}\right\|^2+\left\|\boldsymbol{\sigma}_{\boldsymbol{\theta}}\right\|^2)-\sum_s \log (\frac{\sigma_{s}}{\bar{\sigma}})+\text { const. }
\end{aligned}
\end{equation}

 \subsubsection{Score-based Diffusion Model }
Diffusion models learn the implicit prior of the underlying data distribution by matching its score function ($\nabla \log_{\boldsymbol x} p(\boldsymbol{x})$). 
This prior can be employed for reconstructing images from  undersampled measurements through a forward measurement operator and  detector noise. Let $p(\boldsymbol{x})$ denote an unknown distribution containing samples that follow an independent and identically distributed pattern.
Since calculating 
$\nabla \log_{\boldsymbol x} p(\boldsymbol{x})$ directly is challenging,  the score function can be approximated by a neural network $\boldsymbol{s_{\phi}}$ using  the denoising score matching method \cite{vincent2011connection}. Then the samples belonging to $p(\boldsymbol{x})$ can be obtained via Langevin  MCMC sampling according to the following formula:
\begin{equation}
\boldsymbol{x}_{i+1}=\boldsymbol{x}_{i}+\frac{\eta_i}{2} \boldsymbol{s}_{\boldsymbol{\phi}}\left(\boldsymbol{x}_i\right)+\sqrt{\eta_i} \boldsymbol{z}_i
\end{equation}
where  $\eta_i>0$ represents the step size, and $\boldsymbol{z}_i \sim \mathcal{N} \left(\boldsymbol {0},\boldsymbol{I}\right)$.

\begin{figure*}[!htbp]
\centering{\includegraphics[width=1.9\columnwidth]{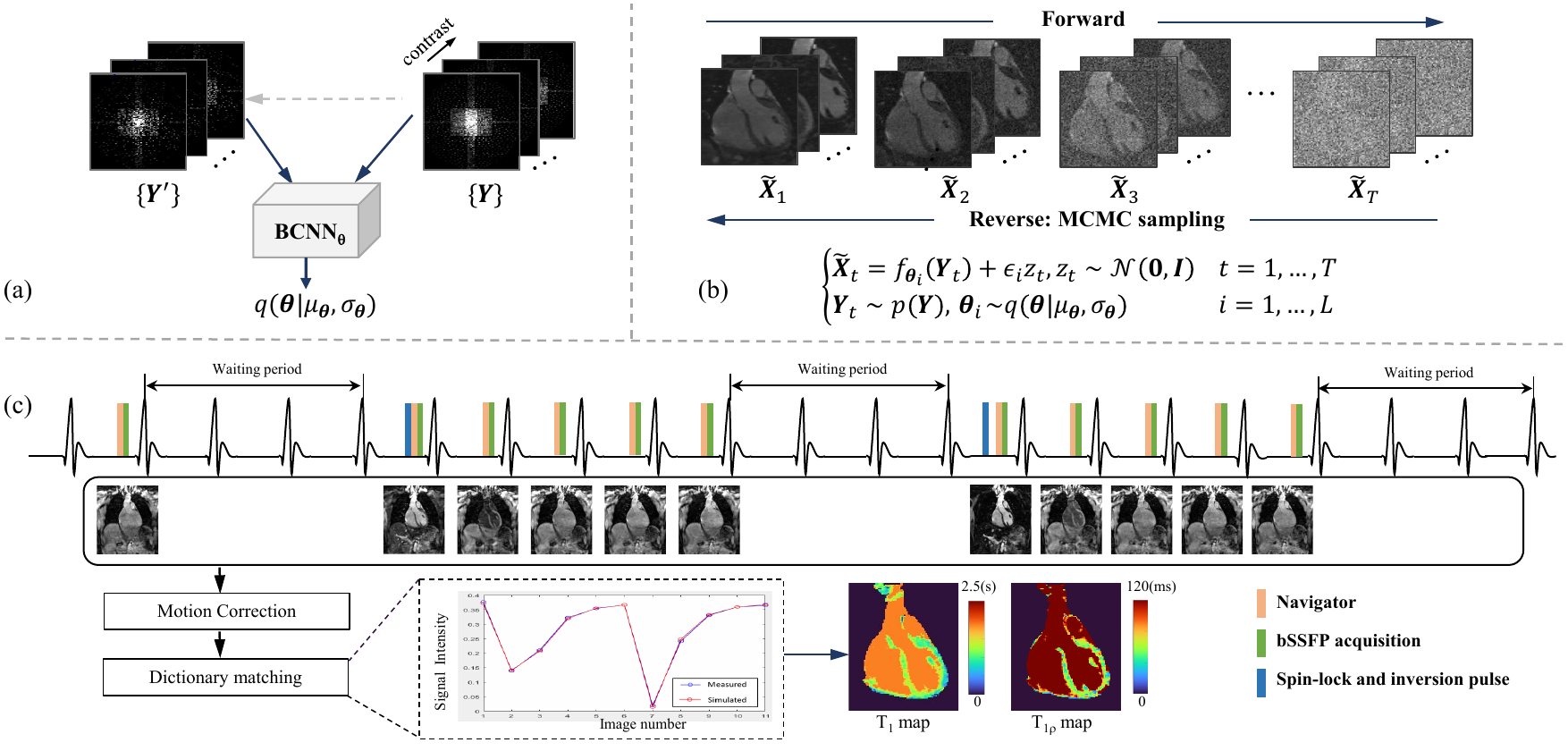}}
\caption{
Illustration of the proposed approach.
(a) Flowchart illustrating the self-supervised BCNN for modeling the parameter distribution.  BCNN takes undersampled $k$-space data pairs $\left\{Y,  Y^{\prime}\right\}$ as input, and produces  $q\left(\theta | \mu_\theta, \sigma_\theta\right)$ as an approximation of $q\left(\theta \mid \mathcal{D} \right)$. 
(b) Forward and reverse diffusion  processes of the score-based model.  In the forward process, the score function, approximating the probability density gradient of  $X$, is learned by perturbing data $\boldsymbol{f}_{\boldsymbol{\theta}}(\boldsymbol{X})$ with multiple noise scales. In the reverse process, multi-contrast images are reconstructed by generating samples from the probability density $p(\boldsymbol X)$ using Langevin MCMC sampling conditioned on the measurement $Y$. (c) Diagram of the pulse sequence and post-processing steps used to estimate $\text T_{1}$ and $\text T_{1\rho}$ maps  via the dictionary matching technique.
}
\label{fig1}
\end{figure*}
\subsection{The Proposed Method}
\subsubsection{MRI Reconstruction Model}
Let  $\boldsymbol X \in \mathbb{C}^{N_{\textit{voxel}} \times N_{\textit {mc}}}$ denote the multi-contrast image series to be reconstructed, where each $\boldsymbol{x}_j$ ($j=1,2,\ldots, N_{\textit{mc}}$) contains $\textit{N}_{\textit{voxel}}$ voxels. 
$\textit{N}_\textit{mc}$ represents the number of contrasts. 
$\boldsymbol Y \in \mathbb{C}^{N_\textit{voxel} \times N_c \times N_{\textit {mc}}}$ is the $k$-space data, where $\boldsymbol{y}_{j}$ is the measurement for each contrast image,    and $N_c$ denotes the number of coils. The reconstruction model to recover $ \boldsymbol{X}$ from its undersampled measurements is typically defined as follows: 


\begin{equation}
\label{MRI forward model}
  \begin{gathered}
\underset{\boldsymbol{X}}{\operatorname{\arg \min}} \frac{1}{2}\left \| \boldsymbol{AX}-\boldsymbol{Y} \right\|_{F}^{2} +\lambda  R(\boldsymbol{X}) 
\end{gathered},  
\end{equation}
where $\boldsymbol A$ represents the encoding operator given by $\boldsymbol A=\boldsymbol{MFS}$,  $\boldsymbol M$ is the undersampling operator that undersamples $k$-space data for each contrast image. $\boldsymbol F$ is the Fourier transform operator, and $\boldsymbol S$ denotes an operator which multiplies the coil sensitivity map with each contrast image coil-wise. $\left\| \cdot \right\|_{F}$ is the Frobenius norm. $R(\boldsymbol{X})$ denotes a combination of regularizations, and $\lambda$ is the regularization weight. In conventional compressed sensing (CS) methods, $R(\boldsymbol{X})$ in  (\ref{MRI forward model}) denotes a combination of sparse\cite{lustig2007sparse}, low-rank\cite{zhang2015accelerating}, or other-type of regularization functions. From a Bayesian perspective,  $R(\boldsymbol{X})$  can be considered as the prior model of the data, denoted as $p(\boldsymbol{X})$. Consequently, it's reasonable to assume that a more precise estimation of this intricate prior data distribution would result in higher-quality samples.
\par Given that fully sampled $\boldsymbol{X}$ is inaccessible, we are compelled to explore a label-free method for estimating $p(\boldsymbol{X})$. To achieve this, a mapping from $\boldsymbol{Y}$ to $\boldsymbol{X}$ is firstly estimated, and then $p(\boldsymbol{X})$ is approximated to characterize the correlation among the MC images.

\subsubsection{Self-supervised Bayesian Reconstruction Network }
\par  Let $\boldsymbol{Y}$ and $\boldsymbol{X}$  denote the undersampled $k$-space data and the corresponding MR image, respectively. A secondary sub-sampling mask is applied to the measurements, resulting in ${\boldsymbol{Y}^{\prime}} = {\boldsymbol{M}^{\prime}} \boldsymbol{Y}$. 
A BCNN can be trained using the sub-sampled data ${\boldsymbol{Y}^{\prime}}$ and the zero-filling image ${\boldsymbol{X}^{\prime}}$ of $\boldsymbol{Y}$ to embody the representation $\boldsymbol{f_{\theta}}$ with network parameter $\boldsymbol{\theta}$:
\begin{equation}
\label{assumption 1}
{\boldsymbol{X}^{\prime}}   = \boldsymbol{f}_{\boldsymbol{\theta}}( {\boldsymbol{Y}^{\prime}} ) +\boldsymbol{n}_1 
\end{equation}
Then the established transformation is utilized to convert the measured ${\boldsymbol{Y}}$ to its corresponding image ${\boldsymbol{X}}$:

\begin{equation}
\label{assumption 1_1}
\boldsymbol{X}= \boldsymbol{f}_{\boldsymbol{\theta}}(\boldsymbol{Y}) +\boldsymbol{n}_2 
\end{equation}
where $\boldsymbol{n}_1$ and $\boldsymbol{n}_2$ represent Gaussian noise with scales $\gamma_1$ and $\gamma_2$. To train the BCNN, we first initialize the parameter $\boldsymbol{\theta}$ 
from a normal Gaussian noise distribution $\boldsymbol{\epsilon}  $$\sim$$ \prod_s \mathcal{N}\left(\epsilon_s | 0,1\right) $, where $\epsilon_s$ denotes the $s$-th element of  $\boldsymbol{\epsilon} $. 
Subsequently, $\boldsymbol{\theta}$ can be sampled through the reparameterization, i.e. $\boldsymbol{\theta}=\mu_{\boldsymbol{\theta}}+\sigma_{\boldsymbol{\theta}} \boldsymbol{\epsilon}$.
Assuming that the measurement noise
$\boldsymbol{n}_1$ follows Gaussian white noise, i.e., $p(\boldsymbol{n}_1) $$\sim$$ \prod_{t } \exp \left(\frac{-{n}_t^2}{2 {\gamma_1}^2}\right)$, and that the prior $p(\boldsymbol{\theta}) $$\sim$$ \prod_s \exp \left(\frac{-{\theta}_s^2}{2 {\bar{\sigma}}^2}\right)$,
the minimum of the KL divergence  in (\ref{loss functon1}) can be calculated by:

\par 

\begin{equation}
\label{loss_BCNN}
\begin{aligned}
& \min _{\boldsymbol{\mu_{\theta}}, \boldsymbol{\sigma_{\theta}}} \frac{1}{2{\gamma_1}^2}\mathbb{E}_{q(\boldsymbol{\theta} \mid \boldsymbol{\mu_{\theta}}, \boldsymbol{\sigma_{\theta}})} \sum_{i=1}^N \left\|\boldsymbol{f_{\theta}}({\boldsymbol{X}^{\prime}}_i) -\boldsymbol{Y}_i\right\|_2^2\\
&+\frac{1}{2\bar{\sigma}^2}  \left(\|\boldsymbol{\mu_{\theta}}\|_2^2+\|\boldsymbol{\sigma_{\theta}}\|_2^2\right)- \sum_s \log \frac{\sigma_{s}}{\bar \sigma} +\text {const.}
\end{aligned}
\end{equation}
where $N$ represents the number of training datasets.

\subsubsection{MC Image Reconstruction via Score-based Model With Joint Distribution }
Utilizing the above estimated $q(\boldsymbol{\theta} )$, we can derive the distribution $p (\boldsymbol {X}|\mathcal{D})$ using the Bayesian equation:

\begin{equation}
\label{bayesion_score}
p(\boldsymbol{X}|\mathcal{D})=\int p(\boldsymbol{X} \mid \boldsymbol{Y}, \boldsymbol{\theta}) \prod_{j=1}^{N_\textit {mc}} p(\boldsymbol{y}_j) q(\boldsymbol{\theta}) \mathrm{d} \boldsymbol{Y} \mathrm{d} \boldsymbol{\theta}
\end{equation}
Since evaluating the integral in (\ref{bayesion_score}) is computationally complex, we employ the score matching technique to approximate $\nabla \log_{\boldsymbol X} p(\boldsymbol{X})$. The fundamental concept behind score matching \cite{hyvarinen2005estimation} is to optimize the parameters $\boldsymbol{\phi}$ of the score-matching network in such a way that 
$\boldsymbol{s_\phi}(\boldsymbol{X};\boldsymbol \theta)=\nabla \log q(\boldsymbol{X};\boldsymbol \theta)$ closely  matches the corresponding score of the true distribution, namely $\nabla \log p(\boldsymbol{X})$. The objective function to minimize is the expected squared error between these two, expressed as: 
$\mathbb{E}_{p(\boldsymbol{X})}\left[1 / 2\left\|\boldsymbol{s_\phi}(\boldsymbol{X})-\nabla \log p(\boldsymbol{X})\right\|^2\right]$. Importantly, this objective is equivalent to the following denoising score matching objective:
\begin{equation}
\label{score function 1}
\min _{\boldsymbol \phi} \mathbb{E}_{p(\boldsymbol{X}, \boldsymbol{Y}, \boldsymbol{\theta})}\left[\frac{1}{2}\left\|\boldsymbol{s_{\phi}}(\boldsymbol{X})-\frac{\partial \log p(\boldsymbol{X} \mid \boldsymbol{\theta}, \boldsymbol{Y})}{\partial \boldsymbol{X}}\right\|^2\right]
\end{equation}


The fundamental concept here is that the gradient $\boldsymbol{s_{\phi}} (\boldsymbol{X})$ of the log density at different corrupted  points  $\tilde{\boldsymbol X}$ should ideally guide us toward the clean sample $\boldsymbol X$. More precisely, we perturb $\boldsymbol{f_{\theta}(Y)}$ by Gaussion noise with different scales $\left\{\varepsilon_i\right\}_{i=1}^L$   which satisfies  $\varepsilon_1<\varepsilon_2<\cdots<\varepsilon_L$. Let $p_{\varepsilon_i}(\tilde{\boldsymbol{X}} |\boldsymbol{Y}, \boldsymbol{\theta})=\mathcal{N}(\tilde{\boldsymbol{X}} | \boldsymbol{f}_{\boldsymbol{\theta}}(\boldsymbol{Y}), \varepsilon_i^2 \boldsymbol{I})$, and the corrupted data distribution can be expressed as $p_{\varepsilon_i}(\tilde{\boldsymbol{X}})=\int p_{\varepsilon_i}(\tilde{\boldsymbol{X}} $$\mid$$ \boldsymbol{Y}, \boldsymbol{\theta}) p(\boldsymbol{Y}) q(\boldsymbol{\theta}) \mathrm{d} \boldsymbol{Y} \mathrm{d} \boldsymbol{\theta}$. It can be observed that $p_{\varepsilon_1}(\boldsymbol{X})=p(\boldsymbol{X})$ when $\varepsilon_1=\gamma_2$. We can estimate the scores of all corrupted data  distributions 
$\{\varepsilon_i\}_{i=1}^L : \mathbf{s}_{\boldsymbol \phi}(\boldsymbol{X},\varepsilon_i) \approx \nabla \log p_{\varepsilon_i}(\tilde{\boldsymbol{X}})$. This is achieved by training the joint score function with the following objective:
\begin{equation}
\label{score_loss2}
\frac{1}{2 L} \sum_{i=1}^L \mathbb{E}_{p(\boldsymbol{Y}) q(\boldsymbol{\theta})} \mathbb{E}_{p_{\varepsilon_i}(\tilde{\boldsymbol{X}} | \boldsymbol{Y},\boldsymbol{\theta})}\left [\left\| \varepsilon_i \boldsymbol{s}_{\boldsymbol{\phi}}(\tilde{\boldsymbol{X}}, \varepsilon_i)+\frac{\tilde{\boldsymbol{X}}-\boldsymbol{f}_{\boldsymbol{\theta}}(\boldsymbol{Y})}{\varepsilon_i}\right\| ^2\right]
\end{equation}

After the score function is estimated, the Langevin MCMC sampling can be applied to reconstruct MC images via the following formula:
\begin{equation}
\label{Langevin sampling}
\begin{aligned}
\tilde{\boldsymbol{X}}_{i+1} & =\tilde{\boldsymbol{X}}_i+\frac{\eta_i}{2} \nabla \log p(\boldsymbol{\tilde{X}}_i | \boldsymbol{Y})+\sqrt{\eta_i} \boldsymbol{z}_i \\
& =\tilde{\boldsymbol{X}}_i+\frac{\eta_i}{2}(\nabla \log p(\boldsymbol{\tilde{X}}_i)+\nabla \log p(\boldsymbol{Y}| \tilde{\boldsymbol{X}}_i))+\sqrt{\eta_i} \boldsymbol{z}_i \\
& =\tilde{\boldsymbol{X}}_i+\frac{\eta_i}{2}(\boldsymbol{s}_\phi(\tilde{{\boldsymbol X}}_i, \varepsilon_i)+\frac{\boldsymbol{A}^H(\boldsymbol{A} \tilde {\boldsymbol{X}}_i-\boldsymbol{Y})}{{\gamma_2}^2+\varepsilon_i^2})+\sqrt{\eta_i} \boldsymbol{z}_i
\end{aligned}
\end{equation}
where $\eta_i$ serves as the step size. 
 The specific values of $\eta_i$ used in this study adhere to those outlined in\cite{song2020improved}, and the complete sampling procedure is shown in Algorithm 1.


\begin{algorithm}[htb]
\caption{Conditional Langevin MCMC Sampling.}
\label{Algorithm 1}
\begin{algorithmic}[1]
\State {\bfseries Input:} $\left\{\varepsilon_i\right\}_{i=1}^L, \epsilon$ and $T$;
\State {\bfseries Initialize:} $\tilde{\boldsymbol{X}}_0 ;$ 
\For{$i=1,\cdots,L$}
\State $ \eta_i=\epsilon \cdot \varepsilon_i^2 / \varepsilon_L^2$ 

\For{$t=1, 2,\cdots, T$}
 \State $\operatorname{Draw} \boldsymbol{z}_t \sim \mathcal{N}(\mathbf{0}, \boldsymbol{I}):$ 
 \State
    $ \tilde{\boldsymbol{X}}_{t}  =\tilde{\boldsymbol{X}}_{t-1}+\frac{\eta_i}{2}(\boldsymbol{s}_\phi(\tilde{\boldsymbol{X}}_{t-1}, \varepsilon_i)+\frac{\boldsymbol{A}^H(\boldsymbol{A} \tilde{\boldsymbol{X}}_{t-1}-\boldsymbol{Y})}{{\gamma_2}^2+\varepsilon_i^2})
  +\sqrt{\eta_i} \boldsymbol{z}_{t} $ 
 
\EndFor
 \State $\tilde{\boldsymbol{X}}_0=\tilde{\boldsymbol{X}}_T$ 
\EndFor
\State {\bfseries Output:} $\tilde{\boldsymbol{X}}_T$.
\end{algorithmic}
\end{algorithm}

 \begin{figure*}[!htbp]
\centering{\includegraphics[width=1.6\columnwidth]{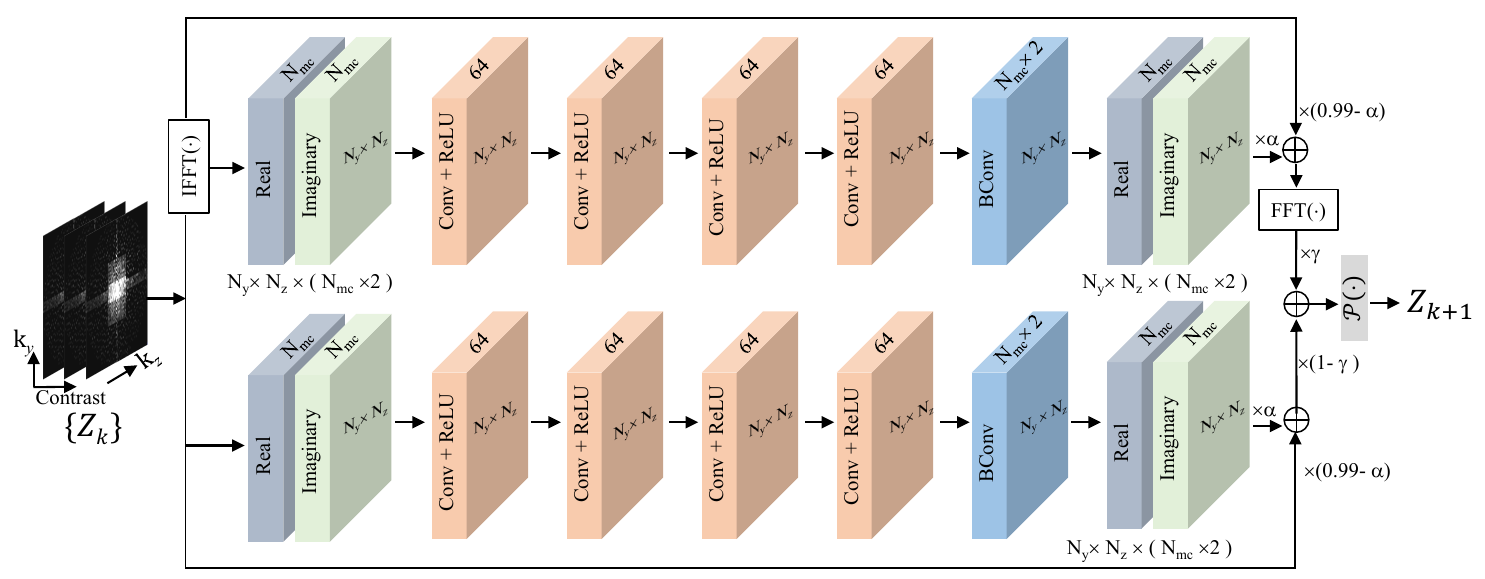}}
\caption{  The network Structure of one block in $f_{\theta}$ used in the proposed method. The operators FFT($\cdot$) and IFFT($\cdot$) represent the Fourier and inverse Fourier transforms, respectively. $\mathcal{P}(\cdot)=(\boldsymbol I-{\boldsymbol{M}^{\prime}})(\cdot)+{\boldsymbol Y^{\prime}}$ denotes projection onto $\boldsymbol Y^{\prime}$. The number above each layer represents the number of output channels, and all convolutional kernels are $3\times 3$.} 

\label{fig2}
\end{figure*}

\subsection{Implementation Details}
Figure 1 illustrates the SSJDM framework. This framework comprises three main components: (a) the BCNN network, (b) the score matching network, and (c) the  joint  $\text T_1$ and $\text T_{1\rho}$ mapping sequence and the parametric maps estimation procedure. Since the MRI data is complex, both real and imaginary components of the multi-contrast data are fed into the BCNN and the score matching networks. In this section, we focus on the implementation specifics of the BCNN and the score matching network\cite{goodfellow2016deep}. 


\par The BCNN network comprises 10 blocks. Each block (as shown in Figure \ref{fig2}) consists of both an upper and a lower module. The upper module exploits redundancy in the image domain, while the lower module ensures inherent self-consistency in the  $k$-space.  Each upper and lower module consists of four convolutional layers and a Bayesian convolutional layer, with each convolutional layer followed by a ReLU activation function. The final layer in both the upper and lower modules outputs the real and imaginary components of the multi-contrast data. To prevent gradient vanishing, we implemented a residual architecture in each module, using the  paradigm of $(0.99-\alpha)\cdot \mathcal{I}+\alpha \cdot \text {CNN}(\cdot)$,$(0\geq \alpha \geq 0.99)$ \cite{cui2023equilibrated}, where $\text {CNN}(\cdot)$ represents the cascaded convolutional neural network part in each module. Notably,  $\boldsymbol{f_{\theta}}$ and the distribution of $\boldsymbol \theta$ together form the BCNN.


\par We used NCSNv2 framework in \cite{song2020improved} as the score matching network. The network input and output have $N_{\textit{mc}} \times 2$  channels.
To establish the joint distribution, the MC images were reshaped so that the input data size of network is $N_b\times (N_{\textit{mc}}  \times 2)\times N_y \times N_z $, where $N_b$ denotes the batch size, $N_z$ and $N_y$ denote the number of the partition and phase encoding lines, respectively. 



 
\section{Experiments}
\subsection{Data Acquisition}
We implemented a 3D simultaneous cardiac $\text T_1$ and $\text T_{1\rho}$ mapping sequence on a 3T MR scanner (uMR 790, United Imaging Healthcare, Shanghai, China). Figure \ref{fig1}(c) shows the timing diagram of the sequence, which closely resembles the one in \cite{milotta20203d} except the preparation module. Here, the preparation module used is a spin-lock and inversion pulse\cite{yang2023robust}, which tips down the magnetization to the -z axis after the $\text T_{1\rho}$ preparation. Then the acquired signals during magnetization recovery contain both $\text T_{1}$ and $\text T_{1\rho}$ weightings. The sequence consists of three parts: (1) The first part acquires an image without the preparation module. (2) In the second part, the preparation module is applied with a spin-lock time (TSL) of 30 ms and an inversion time (TI) of 46 ms, followed by the acquisition of $k$-space segments for 5 images. (3) The third part has the same procedure as the second one, except TSL = 60 ms and TI = 146 ms. At the end of each part, three heartbeats are skipped for magnetization recovery. 

\par Images were acquired using bSSFP readout with ECG trigger and a 2D image navigator (iNAV) for respiratory motion correction \cite{henningsson2012whole,milotta20203d}. The iNAV was obtained using 14 ramp-up preparation pulses (duration = 46 ms) of the bSSFP in each heartbeat, which was used to correct translational motion caused by respiration before image reconstruction \cite{milotta20203d}. 51 volunteers were scanned using 12-channel cardiac and 48-channel spine coils. All cardiac channels were activated, while spine coil channels were manually selected based on the subject's position. This led to a variation in the coil number utilized among subjects, ranging from 20 to 24. Imaging parameters were: field of view (FOV) = 300 × 300 × 96 $\text {mm}^3$, in-plane resolution = 1.7 × 1.7 $\text {mm}^2$, slice thickness = 1.7 mm, slice oversampling = 30\%, TR/TE = 3.28/1.64 ms, segment number = 35, and 11 images was obtained for each acquisition.

\par Two 3D-MC-CMR datasets, namely the training and testing datasets, were acquired using the above sequence. The training dataset consists of prospectively undersampled $k$-space data with R = 6 from 46 volunteers. The acquisition time for each data was approximately 18–19 minutes, depending on the volunteer’s heart rate. The testing dataset includes five prospectively undersampled $k$-space data with R = 2.8, with an average acquisition time of 36.05 minutes. Each testing data was retrospectively undersampled at net acceleration rates of R = 6, 11, and 14, respectively. All undersampling masks used variable-density Poisson-disc sampling scheme, with 24 × 24 $k$-space center fully sampled for coil sensitivity estimation.

\par Each 3D $k$-space dataset was split into 2D slices along the frequency encoding direction by applying the inverse Fourier transform in that direction. The first 30 and last 20 slices were excluded due to the absence of coverage in the cardiac region. A total of 5,796 slices were used for training and 630 slices were used for testing, with 126 slices per subject. The coil sensitivity maps were estimated from the fully sampled $k$-space center using ESPIRiT\cite{uecker2014espirit}, with a kernel size of $6 \times 6$, as well as thresholds of 0.02 and 0.95 for calibration-matrix and eigenvalue decomposition. 

\begin{figure*}[!htbp]
\centering{\includegraphics[width=1.55\columnwidth]{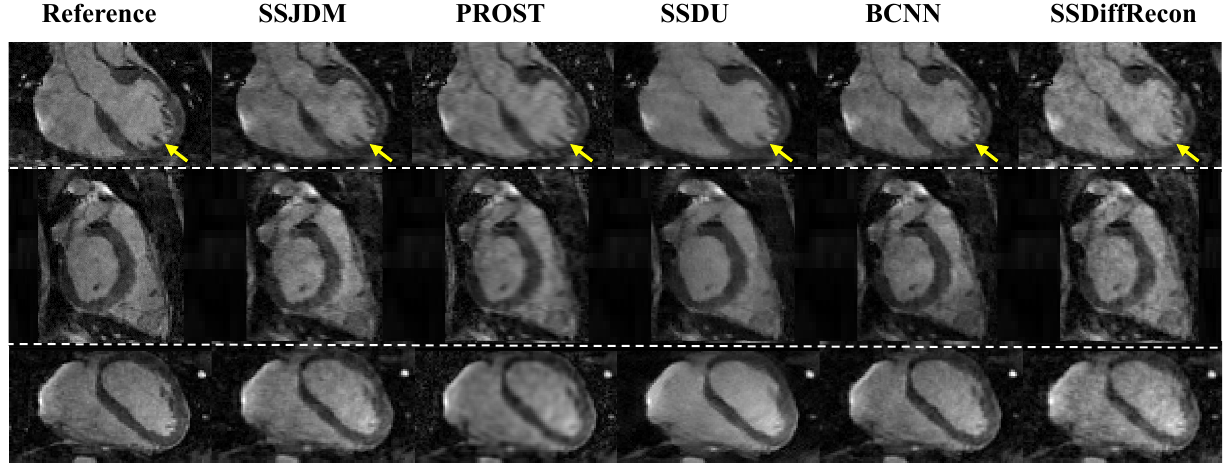}}
\caption{ Reconstructed images at TSL = 30 ms and TI = 46 ms using the SSJDM, PROST, SSDU, BCNN, and SSDiffRecon methods with an acceleration rate of R = 6. The references are from reconstructions using the  CS-based method with R = 2.8. Images of PROST exhibit noticeable blurring, and images of SSJDM show sharper papillary muscles than those of SSDU, BCNN and SSDiffRecon (indicated by the yellow arrows). }
\label{fig3}
\end{figure*}

 \subsection{Reconstruction Experiments} To evaluate the proposed SSJDM method, 
 it was applied to reconstruct the retrospectively undersampled testing data, followed by the estimation of $\text T_1$ and $\text T_{1\rho}$ maps, as described later.  The comparison methods include a traditional regularized reconstruction method (PROST\cite{bustin2019five}), a self-supervised DL method (SSDU \cite{yaman2020self}), BCNN, and a self-supervised reconstruction method with unrolled diffusion models (SSDiffRecon \cite{Yilmaz2023SSDiff}). The images reconstructed from the testing data with R = 2.8, using an ${l_1}$-wavelet regularized compressed sensing method implemented in the BART toolbox \cite{uecker2016bart}, served as the reference.


 \begin{figure*}[!htbp]
\centering{\includegraphics[width=1.55\columnwidth]{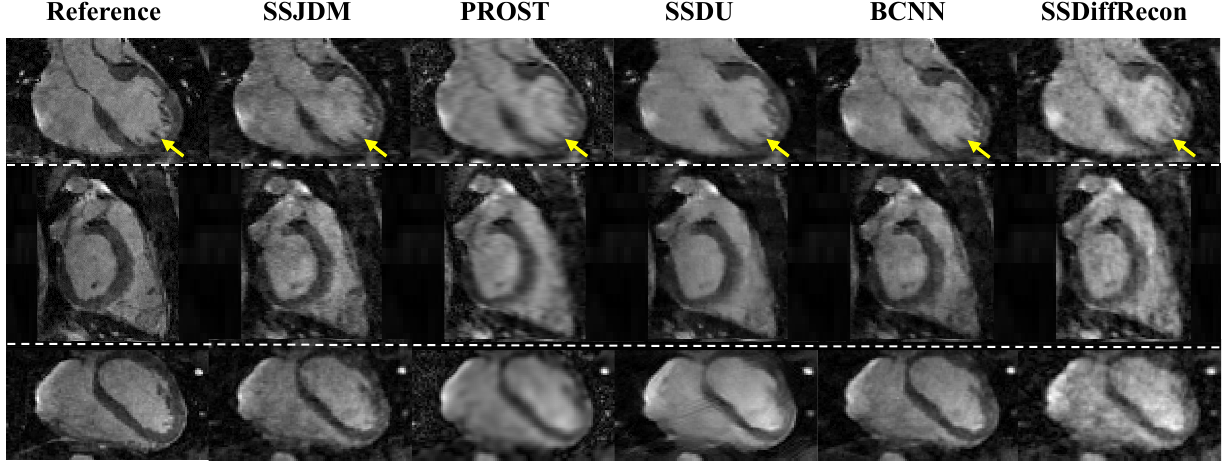}}
\caption{ Reconstructed images at TSL = 30 ms and TI = 46 ms using the SSJDM, PROST, SSDU, BCNN, and SSDiffRecon methods  with an acceleration rate R = 11. The references are from reconstructions using the CS-based method with R = 2.8. Blurring artifacts of PROST  become evident. BCNN and SSDiffRecon preserve more detailed information than SSDU (i.e., the papillary muscle area indicated by the yellow arrows). Images of SSJDM still exhibit sharp boundaries and high texture fidelity.
}
\label{fig4}
\end{figure*}

\par For SSJDM, the BCNN network was configured with $\alpha$ = 0.5 and $\gamma$ = 0.5. Following \cite{yaman2020self}, we selected the subsampled mask $M^{\prime}$ using a variable density strategy based on Gaussian random weighting. The score matching network employed parameters $\varepsilon_1=0.01$, $\varepsilon_L=50$, L = 266, T = 4,  and exponential moving averages (EMA) were enabled with an EMA rate of 0.999.
Both networks were trained using the Adam optimizer, with a learning rate of $10^{-4}$, over 500 epochs, optimizing the respective loss functions in (\ref{loss_BCNN}) and (\ref{score_loss2}), with a batch size of 1. For the PROST method, the parameters included a conjugate gradient tolerance of $10^{-5}$, a maximum of 15 CG iterations, a penalty parameter $\mu$ = 0.08, regularization parameter for high-order singular value decomposition of 0.01, and a maximum of 10 alternating direction method of multipliers iterations. The SSDU methods employed a loss mask ratio of 0.4, and used the Adam optimizer with a learning rate of $10^{-5}$, and had a batch size of 1 during 100 epochs of training. 
The SSDiffRecon method employed a loss mask ratio of 0.95, and used the Adam 
optimizer with a learning rate of $10^{-4}$, and had a batch size of 1 during 2469K steps of training.
SSJDM, SSDU, and SSDiffRecon experiments were conductesd using PyTorch 1.10, TensorFlow 2.10, and TensorFlow 1.14 in Python, respectively, on a workstation equipped with an Ubuntu 20.04 operating system, an Intel E5-2640V3 CPU (2.6 GHz, 256 GB memory), and an NVIDIA Tesla A100 GPU with CUDA 11.6 and CUDNN support. PROST reconstruction was implemented in MATLAB (R2021a, MathWorks, Natick, MA).
Additionally, our reconstruction code for the proposed SSJDM is available at GitHub\footnote{https://github.com/YuanyuanLiuSIAT/SSJDM}.

\par Two radiologists (6 and 4 years of CMR experience, respectively)  independently scored the image quality in the reconstructed MC images based on a 5-point ordinal scale, focusing on image blurring, contrast, and clarity, as well as the clinical diagnostic value for cardiac diseases (1, Poor; 2, Fair; 3, Adequate; 4, Good; 5, Excellent.). Additionally, to evaluate the performance of SSJDM, four prospectively undersampled datasets with R = 11 and R = 14 were acquired from two other volunteers, with each volunteer contributing two datasets, one with R = 11 and another with R = 14.


\subsection{$T_1$ and $ T_{1\rho} $ Estimation}
$\text T_1$ and $\text T_{1\rho} $ parametric maps were estimated using the dictionary matching method\cite{ma2013MRF}.  The dictionary was built by simulated MR signal evolutions generated via Bloch simulation with the imaging parameters as well as the predefined ranges of  $\text T_1$, $\text T_2$, and $\text T_{1\rho} $ values. Subject-specific dictionaries were simulated based on recorded R-R intervals and trigger delays for each scan. The dictionary for this study included 410550 atoms, covering combinations of $\text T_1$ in [50:20:700, 700:10:900, 900:5:1300, 1300:50:3000] ms, $\text T_{1\rho}$ in [5:5:20, 20:0.5:60, 60:5:100, 100:10:300] ms\cite{velasco2022MRF}, and $\text T_2$ in [40:0.5:50] ms\cite{milotta20203dwhole,lyu2023free}, with the notation [lower value: step size: upper value]. Image registration was employed to align the myocardial region in the MC images before dictionary matching. Subsequently, $\text T_1$ and $\text T_{1\rho} $ maps were estimated using the pixel-wise dictionary matching.



\subsection{Ablation Studies}
We conducted three ablation studies to assess the impact of the BCNN network and the joint score-based model on reconstruction performance. 
\subsubsection{Ablation Study 1}
To evaluate the impact of BCNN, we trained the score matching network using MC images reconstructed through conventional CS with total variance (TV) regularization\cite{feng2014golden}.
\subsubsection{Ablation Study 2}
Similar to the first ablation study, we also trained the score matching network using images reconstructed with the SSDU network.

\subsubsection{Ablation Study 3}
To investigate the impact of the joint distribution in the score-based model,  we employed a score-based model with the traditional independent distribution for reconstruction, and compared the results with SSJDM.

\begin{figure*}[!htbp]
\centering{\includegraphics[width=1.55\columnwidth]{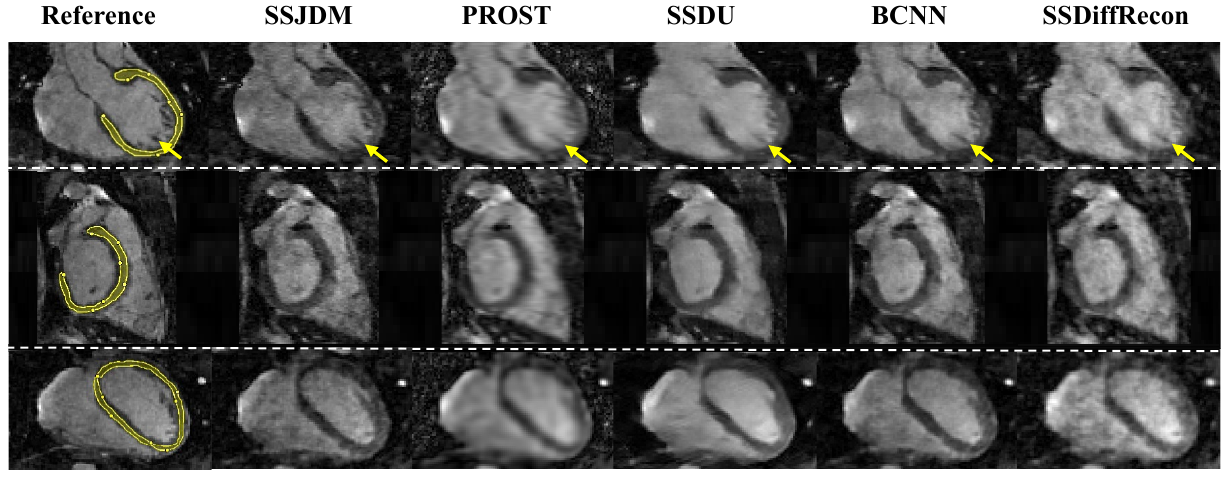}}
\caption{Reconstructed images at TSL = 30 ms and TI = 46 ms using the SSJDM, PROST, SSDU, BCNN, and SSDiffRecon methods with an acceleration rate R = 14. The references are from reconstructions using the CS-based method with R = 2.8. The yellow arrows show the papillary muscle areas in the reconstructed images. Even at a high acceleration rate of R = 14, the image quality of SSJDM degrades a little, while other methods exhibit severe blurring artifacts. The rings with yellow lines in figures of the first column show an illustration of the selected ROIs in the myocardial region used for the $\text {T}_1$  and $\text {T}_{1\rho}$ evaluations.}
\label{fig5}
\end{figure*}

\begin{figure*}[!htbp]
\centering{\includegraphics[width=1.6\columnwidth]{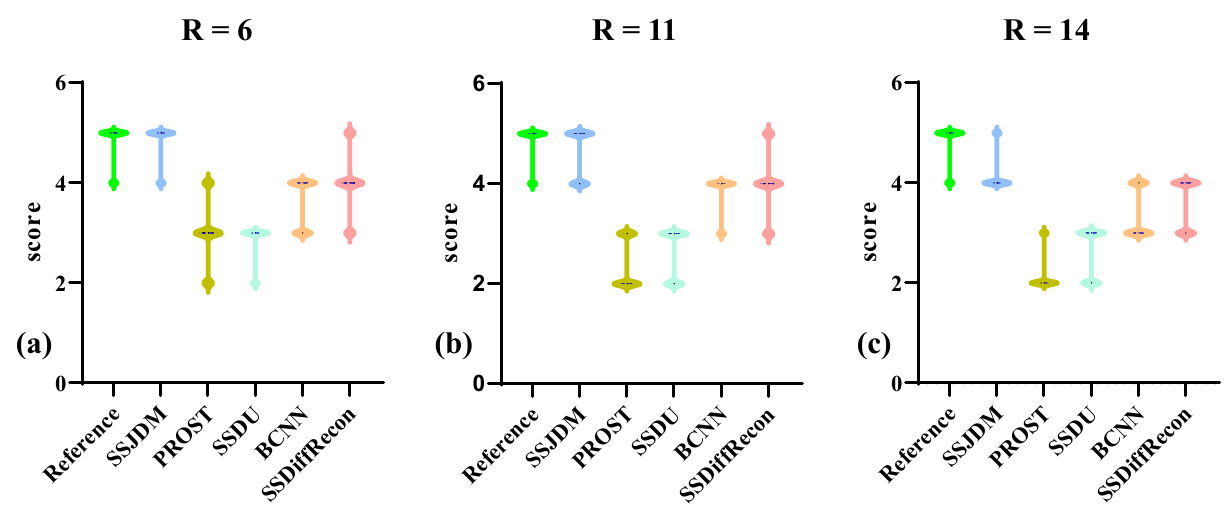}}
\caption{  The image quality evaluation results based on a 5-point scale at different acceleration rates using different reconstruction methods. SSJDM at all rates were evaluated to be significantly improved compared with other four methods in terms of blurring and overall image quality. }
\label{figure_6_violin_plot}
\end{figure*}

\begin{figure}[!htbp]
\centering{\includegraphics[width=0.8\columnwidth]{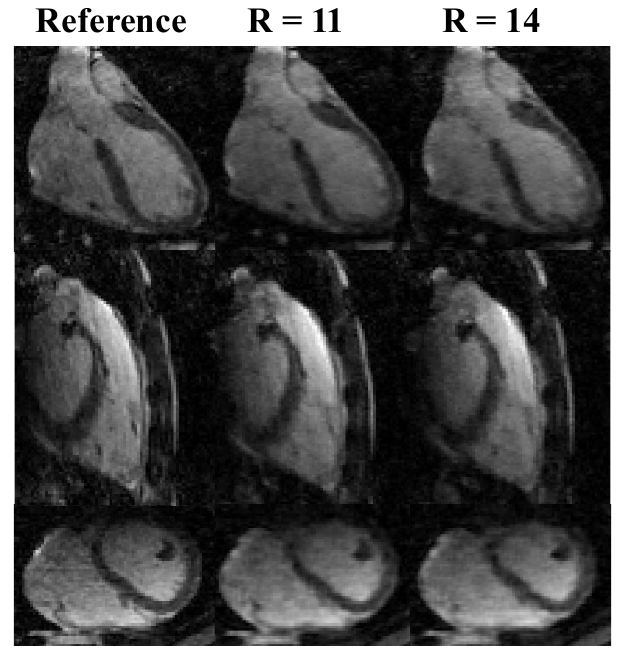}}
\caption{  Images reconstructed using SSJDM reconstructed from prospectively undersampled datasets at R = 11 and 14. The references are from reconstructions using the CS-based method with R = 2.8. }
\label{fig7}
\end{figure}

\begin{figure*}[!htbp]
\centering{\includegraphics[width=2\columnwidth]{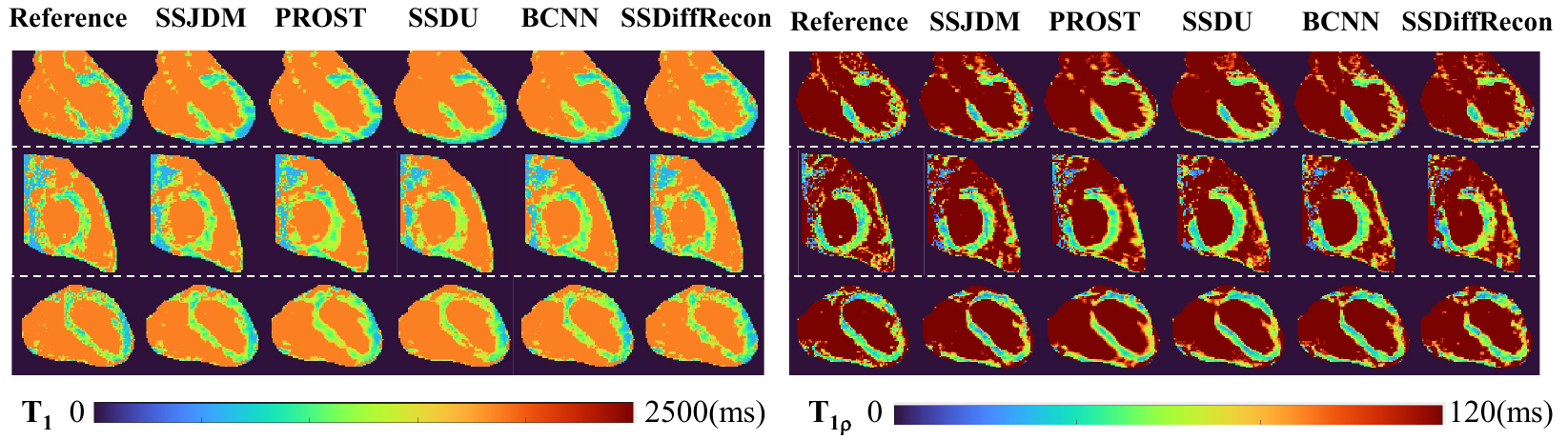}}
\caption{ $\text T_{1}$ and $\text T_{1\rho}$ maps estimatd from reconstructed images using the proposed SSJDM, the PROST, SSDU, BCNN, and SSDiffRecon methods  with acceleration rate R = 11. The reference maps are estimated from the reconstructions using the CS-based method with R = 2.8. The maps of SSJDM exhibit sharp boundaries and high texture fidelity.}
\label{fig8}
\end{figure*}


\section{Results}
Figure \ref{fig3}  shows the first image after the preparation pulse with TSL = 30 ms and TI = 46 ms from one volunteer, reconstructed using the SSJDM, PROST, SSDU, BCNN, and SSDiffRecon methods at R = 6. To focus on the region of interest, the reconstruction images were cropped to display only the heart. Among the reconstructed results, the images of PROST exhibit noticeable blurring.  The rest four methods successfully reconstructed images at this acceleration rate, while the images of SSJDM show sharper papillary muscles than those of SSDU, BCNN, and SSDiffRecon (indicated by the yellow arrows). 
Figure \ref{fig4} shows the reconstructed results using the five methods at R = 11.
Blurring artifacts of PROST become noticeable in scenarios with higher acceleration rates. BCNN and SSDiffRecon preserve more detailed information than SSDU, notably in the reconstructed papillary muscle area (indicated by the yellow arrows), since both methods improved SSDU by capturing data distribution for model training \cite{jospin2022hands}.
Images of SSJDM still exhibit sharp boundaries and high texture fidelity, highlighting the superior performance of score-based reconstruction methods in restoring image details. At an even higher acceleration rate of R = 14  (shown in Figure \ref{fig5}), the image quality of SSJDM degrades a little, while other methods exhibit severe blurring artifacts. Table \ref{table 1} shows the average PSNR, SSIM, and RMSE values for all reconstructed images. The proposed SSJDM method qualitatively achieves the best performance among the five methods across all acceleration factors.
Figure \ref{figure_6_violin_plot} shows the violin plots of the scores for the image quality evaluation study. For all acceleration rates, SSJDM gets the highest scores compared to the other four methods in terms of the blurring and overall image quality.

 Figure \ref{fig7} shows reconstructed images using SSJDM from the prospectively undersampled datasets at R = 11 and 14, which further confirms the effectiveness of the proposed SSJDM method for 3D-MC-CMR reconstruction.

\par Figure \ref{fig8} shows  the $\text T_1$ and $\text T_{1\rho} $ maps estimated from reconstructions using different methods at R = 11. Similar conclusions can be drawn from the estimated maps as those from the reconstructed MC images at the same acceleration rate shown in Figure \ref{fig4}. 
Table \ref{table 2} shows the mean and standard deviation of $\text T_1$, $\text T_{1\rho} $ values using different methods in manually drawn myocardium regions at different acceleration rates and the reference values. The  $\text T_1$ and $\text T_{1\rho}$ values estimated from the SSJDM reconstructions are consistent with the reference values.

\begin{table}[!htbp]
\caption{COMPARISONS OF DIFFERENT METHODS  WITH DIFFERENT ACCELERATION FACTORS (R). THE BEST RESULTS ARE IN BOLD.}
\begin{center}
\setlength{\tabcolsep}{0.60mm}{
\begin{tabular}{cc|ccc}
\hline \hline  & Metrics & PSNR & SSIM & RMSE   \\
\hline \multirow{5}{*}{ 6$\times$}& SSJDM & \textbf{29.659 $\pm$ 1.056 } & \textbf{0.933 $\pm$ 0.002}  & \textbf {0.033 $\pm$ 0.004}  \\
& PROST & {27.096 $\pm$ 1.575}  & 0.910 $\pm$ 0.004  & 0.044 $\pm$ 0.007  \\
& SSDU & {27.533 $\pm$ 2.163 } & 0.917 $\pm$ 0.004  & 0.042 $\pm$ 0.011   \\
&BCNN & {28.273 $\pm$ 1.824 } & 0.922 $\pm$ 0.005  & 0.037 $\pm$ 0.010  \\
&SSDiffRecon & {27.046 $\pm$ 1.995
 } & 0.921 $\pm$ 0.009  & 0.044 $\pm$ 0.018    \\
\hline \multirow{5}{*}{11$\times$}& SSJDM & \textbf{28.049 $\pm$ 1.039 } & \textbf{0.916 $\pm$ 0.003}  & \textbf{0.039 $\pm$ 0.004 }   \\
& PROST & {22.350 $\pm$ 2.044}  & 0.879 $\pm$ 0.004  & 0.076 $\pm$ 0.014  \\
& SSDU & {24.356 $\pm$ 1.917 } & 0.887 $\pm$ 0.005  & 0.064 $\pm$ 0.017   \\
&BCNN & {27.220 $\pm$ 1.283 } & 0.907 $\pm$ 0.005  & 0.049 $\pm$ 0.001   \\
&SSDiffRecon & {25.485 $\pm$ 0.993
 } & 0.902 $\pm$ 0.006  & 0.053 $\pm$ 0.016   \\
\hline \multirow{5}{*}{14$\times$}& SSJDM & \textbf{27.994 $\pm$ 0.598  } & \textbf{0.910 $\pm$ 0.002 } & \textbf{0.040 $\pm$ 0.003}    \\
& PROST & {21.640 $\pm$ 1.485}  & 0.868 $\pm$ 0.002  & 0.083 $\pm$ 0.010 \\
& SSDU & {22.694 $\pm$ 1.213 } & 0.874 $\pm$ 0.003  & 0.080 $\pm$ 0.013   \\
&BCNN & {25.161 $\pm$ 1.005 } & 0.891 $\pm$ 0.003  & 0.055 $\pm$ 0.007   \\
&SSDiffRecon & {23.861 $\pm$ 0.510 } & 0.887 $\pm$ 0.005 & 0.066 $\pm$ 0.006  \\
\hline \hline
\end{tabular}}
\label{table 1}
\end{center}
\end{table}

\par
\begin{figure}[!htbp]
\centering{\includegraphics[width=1\columnwidth]{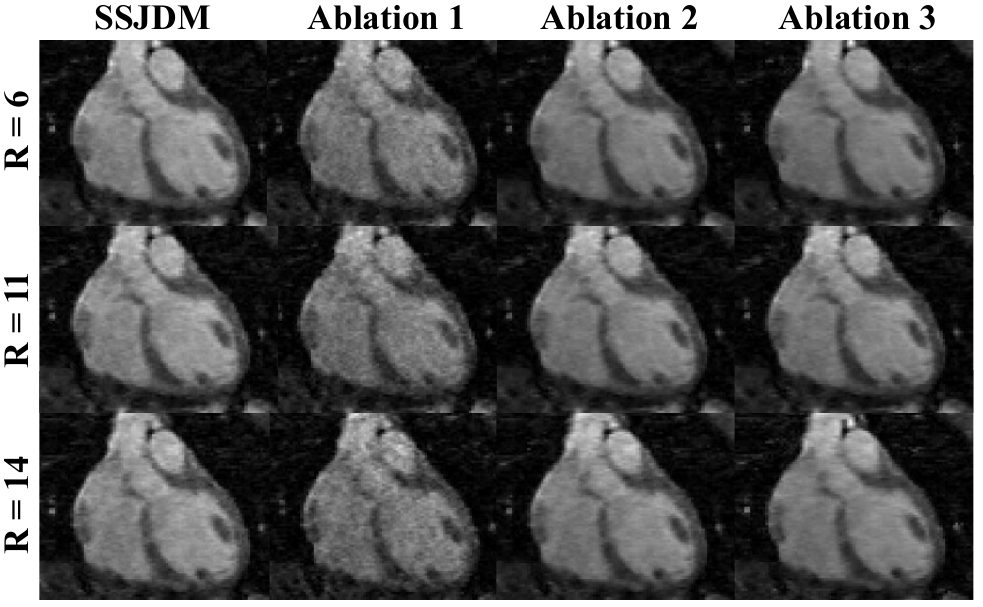}}
\caption{ Reconstructed images at different acceleration rates in the ablation studies. Images of Ablation 1 exhibit severe noise, and images of Ablation 2 and Ablation 3 exhibit blurring compared to those of SSJDM. }
\label{fig9}
\end{figure}

 Figure \ref{fig9} shows reconstructed images at R = 6, 11, and 14 using SSJDM and the methods  from three ablation studies. Ablation 1 exhibits severe noise, attributed to the training data originating from conventional CS, which is prone to substantial noise amplification at higher acceleration rates\cite{yaman2020self}. 
Although noise is effectively mitigated by substituting the input data with reconstructions from the SSDU method, the resulting images (referred to as "Ablation 2") still display blurriness, especially at higher acceleration rates, such as R = 14. Images reconstructed using the independent distribution (referred to as "Ablation 3") exhibit slight blurring compared to SSJDM, indicating an enhanced reconstruction quality with the joint distribution.



\begin{table}[!htbp]
\caption{THE ESTIMATED MYOCARDIAL $\text T_1$ AND $\text T_{1\rho}$ VALUES (MEAN $\pm$ STD, UNIT:MS) 
FOR SSJDM, PROST, SSDU, BCNN, AND SSDIFFRECON methods. THE REFERENCE VALUES ARE ALSO PROVIDED FOR COMPARISON. }
\begin{center}
\setlength{\tabcolsep}{0.58mm}{
\begin{tabular}{c|c|ccc}
\hline \hline  
Reference & AF & R = 6 & R = 11 & R = 14   \\
\hline 
\multirow{5}{*} {\makecell{$\text {T}_{1}$ \\ (1037.8$\pm$27.1)}}
&  SSJDM & \textbf{1041.5 $\pm$ 25.3} & \textbf{1044.0 $\pm$ 31.4} & \textbf{1046.9 $\pm$ 31.7}  \\
 &  PROST & {1007.6 $\pm$ 29.8} & {1002.1 $\pm$ 34.9} & {1000.9 $\pm$ 36.8}  \\
&  SSDU & {1016.1 $\pm$ 29.4} & {1007.6 $\pm$ 34.1} & {1004.6 $\pm$ 35.7}  \\
 &  BCNN & {1024.0 $\pm$ 25.8} & {1021.3 $\pm$ 34.6} & {1020.0 $\pm$ 35.4}  \\
&  SSDiffRecon & {1022.6 $\pm$ 26.0} & {1020.3 $\pm$ 35.1} & {1013.7 $\pm$ 35.9}  \\
\hline
\multirow{5}{*}
{\makecell{$\text T_{1\rho}$ \\ (56.1 $\pm$ 3.2)}}
&  SSJDM & \textbf{56.8 $\pm$ 4.2
} & \textbf{55.2 $\pm$ 5.1
} & \textbf{54.9 $\pm$ 5.9}  \\
 &  PROST & {59.0 $\pm$ 4.3} & {60.0 $\pm$ 6.0
} & {63.3 $\pm$ 7.0}  \\
&  SSDU & {53.6 $\pm$ 5.0} & {52.8 $\pm$ 5.7} & {52.5 $\pm$ 6.1}  \\
 &  BCNN & {57.2 $\pm$ 4.9} & {58.5 $\pm$ 5.6} & {59.0 $\pm$ 6.0}  \\
&  SSDiffRecon & {62.1 $\pm$ 4.9} & {59.9 $\pm$ 5.9} & {60.5 $\pm$ 6.5}  \\

\hline \hline
\end{tabular}}
\label{table 2}
\end{center}
\end{table}

 \section{Discussion}

In this study, we proposed the SSJDM method, which utilizes a self-supervised BCNN to train the score-based model without the need for high-quality training data. This approach enabled the successful reconstruction of MC images from highly undersampled $k$-space data. We have shown that our method can mitigate the discrepancy between the output of single-model-based methods (i.e., CS-based or generally deterministic networks-based methods) and actual data, since this method considers the stochastic nature of parameters in the reconstruction model. This enhancement leads to improved image quality of the score-based model, resulting in superior performance compared to self-supervised SSDU and BCNN approaches. The SSJDM method demonstrated favorable reconstructions in 3D cardiac imaging. It allowed for the simultaneous estimation of $\text T_1$ and $\text T_{1\rho}$ parametric maps in a single scan. Moreover, the estimated $\text T_1$ and $\text T_{1\rho}$ parameter values exhibited a strong consistency with those obtained from conventional 2D imaging.

\par Acquiring high-quality training samples is super challenging in MRI, especially in 3D cardiac imaging. The fully sampled acquisition for the 3D-MC-CMR imaging sequence in this study would be over 2 hours. This prolonged scan duration could easily induce subject motion, leading to motion artifacts. Therefore, conventional supervised reconstruction techniques are no longer suitable for such scenarios.
Unsupervised learning-based imaging techniques,  including self-supervised, untrained, and unpaired learning approaches, may provide an alternative to tackle this challenge.
Untrained imaging methods, such as traditional CS-based methods \cite{lustig2007sparse}, untrained neural network-based methods like deep image prior (DIP)\cite{ulyanov2018deep}, and $k$-space interpolation with untrained neural network (K-UNN)\cite{cui2023k}, do not rely on training data and have exhibited remarkable performance. However, CS-based methods have computationally intensive processes, and optimizing the regularization parameters is non-trivial. Untrained neural network-based methods rely on the neural network structure for implicit regularization and face challenges in selecting an appropriate regularization parameter and early stopping condition when fitting the network to the given undersampled measurement. Unpaired deep learning approaches based on generative models, such as variational autoencoders\cite{kingma2013auto}, generative adversarial networks\cite{zhu2017unpaired}, normalizing flows\cite{dinh2014nice}, and the original diffusion model\cite{ho2020denoising}, have been introduced for MR image reconstruction. The related networks are typically trained without paired label data.  However, the training process for these approaches still requires  fully sampled datasets, making them different from our study and unsuitable to our application scenario.
Therefore, we undertake the reconstruction of 3D-MC-CMR images in a self-supervised learning manner.

\par In ablation studies,  we observed a degradation in reconstruction quality when training the score matching network with images reconstructed through conventional CS method. As mentioned in previous literatures, higher factors can lead to degradation in CS reconstructions, which in turn impact the performance of the trained score matching network. Employing training datasets generated by a more effective method, such as the SSUD method, can improve the reconstruction performance of the score matching network. However, the improvement is restricted because the SSDU method learns the transportation mapping with deterministic weights. Either CS or SSDU approach entails the use of a single model or a network with fixed parameters to reconstruct training samples for the score matching network. However, these approaches may fall short in adequately capturing the diversity inherent in real data.  The BCNN provides a natural approach to quantify uncertainty in deep learning\cite{jospin2022hands}, i.e., leveraging a set of models or parameters to comprehensively represent the diversity within real data can effectively serve the purpose of ensemble learning. 
In this study, the  proposed SSJDM method learns the transportation mapping built on the framework of Bayesian neural network with the weights treated as random variables following a certain probability distribution. In contrast, in ablation studies 1 and 2, the input data correspond to a learned transportation mapping with deterministic weights. Consequently, this self-supervised BCNN training seems to enhance the network's generalization capacity, and this learning strategy holds potential for further improving the reconstruction performance.

\par This study extends our prior self-supervised research presented in \cite{Cui2022} (referred to as "Self-score") from an application-oriented standpoint, given the absence of fully-sampled acquisitions in 3D-MC-CMR imaging. The SSJDM method differs from the Self-score approach by capturing the joint distribution of MC images in training the score-based model, while the Self-score method trained the model with an independent distribution. The SSJDM method achieves enhanced reconstructions, as demonstrated in the ablation study. The proposed reconstruction framework also demonstrates excellent performance in practical applications.
\begin{table}[htbp]
\caption{THE NUMBER OF MODEL PARAMETERS, TRAINING AND AVERAGE INFERENCE TIMESFOR ALL COMPARISON METHODS}
\begin{center}
\setlength{\tabcolsep}{0.58mm}{
\begin{tabular}{c|ccccc}
\hline \hline
Methods& Parameters & Training time (h) & Inference time (s) \\ 
\hline
SSJDM (GPU) & 59475756 & 250.02 & 3010.38 \\ 
PROST (CPU) & / & / & 16719.04 \\ 
SSDU (GPU) & 592129 & 132.72 & 939.32 \\ 
BCNN (GPU) & 29737878 & 127.35 & 1513.96 \\ 
SSDiffRecon (GPU) & 732570 & 140.45 & 945.01 \\ 
\hline \hline
\end{tabular}}
\label{table 3}
\end{center}
\end{table}
\par There are several limitations in this study. First, the inference time is relatively long. As shown in Table \ref{table 3}, the training and average inference times, and number of model parameters are provided for all comparison methods. The SSJDM method consumes much longer inference time than end-to-end DL-based reconstruction methods. This limitation arises from a common drawback of score-based generative models, which require large sampling steps of the learned diffusion process to achieve the desired accuracy. Specifically, the SSJDM method captures correlations between multi-contrast images through joint distribution modeling, whereas some existing diffusion-based methods employ low-rank constraints that rely on singular value decomposition (SVD), making them more time-consuming. Consequently, SSJDM shows improved computational efficiency relative to these diffusion models with low-rank constraint. To reduce inference time, various sampling acceleration strategies have been proposed, such as consistency models\cite{song2023consistency}, ordinary differential equation solver\cite{lu2022dpm}, demonstrating that similar results can be obtained in fewer than ten steps. However, most of their performance has not yet been evaluated in MR reconstruction. We aim to explore these sampling-acceleration strategies in our future work to reduce the inference time of the SSJDM method, making it more suitable for clinical use. Secondly, dictionary-matching for $\text T_1$ and $\text T_{1\rho}$ estimation is another time-consuming procedure, depending on dictionary size and pixel count. Techniques, like dictionary compression\cite{mcgivney2014svd} and DL network for direct mapping\cite{cohen2018mr, fang2019deep}, offer potential solutions to address this challenge. As this study focuses on reconstructing 3D CMR images, we’ll investigate the fast parameter estimation method in the future.

\section{CONCLUSIONS}
This work shows the feasibility of score-based models for reconstructing 3D-MC-CMR images from highly undersampled data in simultaneous whole-heart $\text T_1$ and $\text T_{1\rho}$ mapping. The proposed method is trained in a self-supervised manner and therefore particularly suited for 3D CMR imaging that lacks fully sampled data. Experimental results showed that this method could achieve a high acceleration rate of up to 14 in simultaneous 3D whole-heart $\text T_1$ and $\text T_{1\rho}$ mapping.

\section*{Supplementary Information }

\setcounter{table}{0}
\setcounter{figure}{0}
\renewcommand\thefigure{S\arabic{figure}}
\renewcommand\thetable{S\arabic{table}} 

\subsection{Motion Correction}
\begin{figure}[!htbp]
\centering{\includegraphics[width=1\columnwidth]{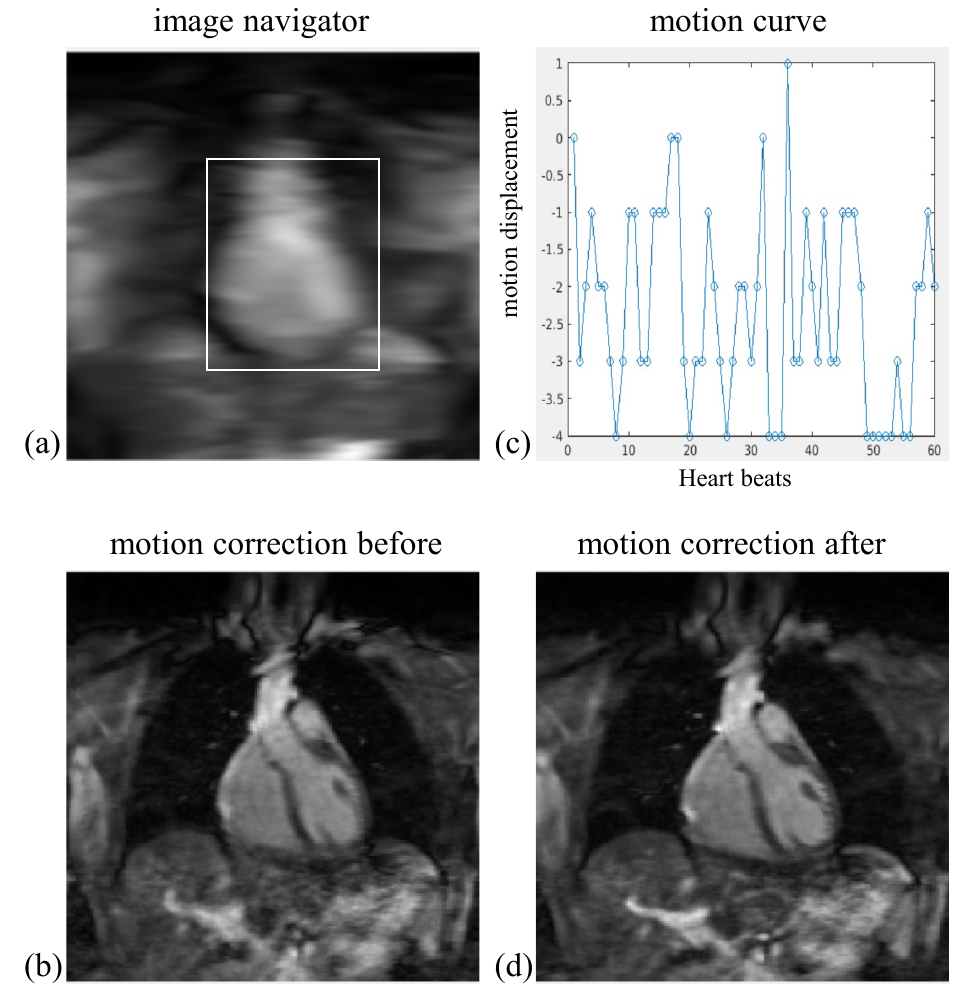}}
\caption{(a) The 2D image navigator (iNAV) (b) The measured motion curve from  iNAV (c) The motion-corrupted zero-filling  image (d) The motion-corrected zero-filling image.}
\label{figure S1}
\end{figure}

The respiratory motion correction process consists of three main steps: 1. Image Acquisition: 2D image navigators (iNAVs) \cite{henningsson2012whole} are acquired prior to each volume by spatially encoding the 14 ramp-up pulses of the bSSFP readout in each heartbeat. These iNAVs are low-resolution 2D images in the coronal plane, as shown in Figure \ref{figure S1}(a). 2. Motion Estimation: The 2D translational motion, including foot-head (FH) and left-right (LR) directions, is estimated using a template-matching algorithm. A template (indicated by the white rectangle in \ref{figure S1}(a))  is manually selected around the heart, and FH and LR beat-to-beat translations are calculated using a mutual information similarity measure. An example of an estimated FH motion curve for an MC image is provided in Figure \ref{figure S1} (b). 3. Motion Correction: Motion correction is applied by modulating the k-space data with a linear phase shift to align it with a reference position at end-expiration. Zero-filling reconstructions of the undersampled 3D-MC-CMR datasets before and after motion correction are shown in Figures \ref{figure S1}(c) and \ref{figure S1}(d), respectively. 

\bibliography{Ref_Bib}
\bibliographystyle{IEEEtran}

\end{document}